\documentstyle[aps,multicol,epsf]{revtex}

\draft

\def\squote{}
\def\quote#1#2#3#4{\squote {#1,\ {\sl#2}\ {\bf#3}, #4}.\par}
\def\qquote#1#2#3#4{\squote {#1,\ {\sl#2}\ {\bf#3}, #4};}

\def\prl{{\sl Phys. Rev. Lett.}\ }

\def\pr {{\sl Phys. Rev.}\ }

\def\w{\omega}
\def\e{\epsilon}
\def\deriv{\partial}
\begin{document}
\title{From the zero-field metal-insulator transition in
 two dimensions to the quantum Hall transition: 
a percolation-effective-medium theory}
\bigskip
\author{\large  Yigal Meir}
\address{Department of Physics, Ben-Gurion University, Beer Sheva 84105, ISRAEL
\\and\\The Ilse Katz Center for Meso- and Nanoscale Science and Technology,\\
 Ben-Gurion University, Beer Sheva 84105, ISRAEL} \maketitle
\begin{abstract}
Effective-medium theory is applied to the percolation description
of the metal-insulator transition in two dimensions with emphasis
on the continuous connection between the zero-magnetic-field
transition and the quantum Hall transition. In this model the system
consists of puddles connected via saddle points, and there is loss of
quantum coherence inside the puddles. The effective conductance of the
network is calculated using appropriate integration over the 
distribution of conductances, leading to
a determination of the magnetic field dependence of the critical density.
Excellent quantitative agreement is 
obtained with the experimental data, which allows an estimate of
the puddle physical parameters. 

\end{abstract}
\begin{multicols}{2}
Extensive experimental and theoretical effort has been invested in
the attempt to understand and characterize the observation of
metallic-like behavior in two dimensions \cite{abrahamsreview}. One of
the  intriguing experimental findings \cite{hanein} is that this
zero-magnetic-field "transition" is continuously connected with
the integer quantum Hall transition. In previous publications\cite{myperc}
I presented a simple non-interacting electron model,
 combining local quantum transport and global classical
percolation,  which treated the zero field transition and the
quantum Hall transition on the same footing. Numerical calculations
showed behavior qualitatively similar to that observed experimentally.
In the present paper I present an analytic approach,  based on the
 effective-medium theory of percolation \cite{kirkpatrick},
   that allows quantitative
comparison with the experiment. This comparison also allows determination
of physical properties of the underlying electronic state. The implications
on transport in the quantum Hall regime due to this continuous connection
 to the zero-field transition are discussed and shown to be in agreement
with experimental observations in the quantum Hall regime.

The model is based on two  assumptions:
(a)  the potential fluctuations due to the
 disorder define density puddles,  connected via saddle points,
  or quantum point-contacts (QPCs),  and (b)
 The electron wavefunction totally dephases in the puddles, i.e. 
 the time the particle spends in the puddle is larger that the dephasing
time. 
 Each saddle point is characterized 
 by its critical energy $\e_c$, such that the
 transmission through it is given by $T(\e)=\Theta(\e-\e_c)$,
 where quantum tunneling has been neglected.
 Thus
 the conductance through each QPC is given by the Landauer formula,
\begin{eqnarray}
G(\mu,T) &=& {2e^2\over h} \int d\e
 \left(-{{\deriv f_{FD}(\e)}\over{\deriv \e}}\right)
 T(\e)\nonumber\\
  &=&  {2e^2\over h} {1\over{1+\exp[(\e_c-\mu)/kT]}} ,
\label{QPC}
\end{eqnarray}
where $\mu$ is the chemical potential,  and
$f_{FD}$ is the Fermi-Dirac distribution function.

The system is now composed of classical resistors,  where the
 resistance of each one
 of them is given by (\ref{QPC}),  with random QPC energies.
Effective medium theory for such random resistor network has been
developed many years ago \cite{kirkpatrick}, and it agrees with exact
solution of this network on an infinite lattice (Cayley tree).
 The resulting equation for the 
effective conductance $\sigma$ of the total network is given by
\begin{equation}
0 = \int dG f(G) {{G-\sigma}\over{G+(z-2)\sigma}} , 
\label{EMTequation}
\end{equation}
where $f(G)$ is the distribution function of the conductances in the
network, and $z$ is the coordination number of the lattice.

In the present case,  the conductance through a point contact,  given
by (\ref{QPC}), depends 
on the the threshold energy of the quantum point contact and the 
chemical potential in the puddles that are connected by it. At zero 
temperature it reduces to  
$G(\mu,T=0) = 2e^2/h\times \Theta(\mu-\e_c)$, where $\Theta(x)$ is the
Heaviside step function.
For an arbitrary distribution
of  the threshold energies $f_{thr}(\e_c)$,  the equation for the effective
conductance of the network reads
\begin{equation}
0 = I_f {{1-\sigma}\over{1+(z-2)\sigma}} +\ (1-I_f) {{-\sigma}\over{(z-2)\sigma}}, 
\label{ecequation}
\end{equation}
with $I_f\equiv\int_{-\infty}^\mu f_{thr}(\e_c) d\e_c$. Solving for 
$\sigma$,  one finds
\begin{equation}
\sigma(\mu) = \sigma_0 \times (\mu-\mu_c) , 
\label{linear}
\end{equation}
with $\sigma_0$ (of the the order of $e^2/h$)
 and $\mu_c$ nonuniversal constants,  depending on the
distribution function and the lattice. The conductance vanishes below 
a critical chemical potential (density) and grows linearly above it.
The mean-field  critical 
conductance exponent was found  to be unity (compared to $\simeq1.3$ in
two dimensions).

In the presence of a finite perpendicular magnetic field $B$ the situation
becomes more complicated. Landau levels form in the puddles (see inset to
Fig.~1),  and the
chemical potential in each puddle oscillates with magnetic field,  as
it is stuck in the highest occupied Landau level in the puddle. 
If a puddle is modeled by potential parabolic well,
 characterized by it depth, $\e_0$, 
and by its confining energy,  $\hbar\w_0$ (which also
defines its characteristic size),  then 
the chemical potential is given by
\begin{equation}
\mu(\mu_0,\e_0,\w_0,B) = \e_0 + \hbar \w/2 +  \hbar \w
\left[ {{\mu_0-\hbar\w_0/2-\e_0}\over{ \hbar \w}}\right]_{int} , 
\label{mu}
\end{equation}
where $[\ \ ]_{int}$ denotes the integer part, 
 $\w \equiv \sqrt{\w_0^2+\w_c^2}$,  and
$w_c(B)$ is  the cyclotron energy. The parameter $\mu_0$ is 
 the zero-field chemical potential (density). For large field
the chemical potential is $\e_0 + \hbar \w/2$, the energy of the
 lowest Landau level. As magnetic field is reduced,
there will be a first jump in the chemical potential for
$\mu_0-\hbar\w_0/2-\e_0\simeq \hbar \w$, which defines filling
factor $\nu=1$.
In Fig.~1 a typical magnetic field
dependence of the chemical potential in two puddles is depicted.
 The two puddles 
differ only in the value of $\e_0$, the depth of the potential. 

\begin{center}
\leavevmode
\epsfxsize=4in
\epsfbox{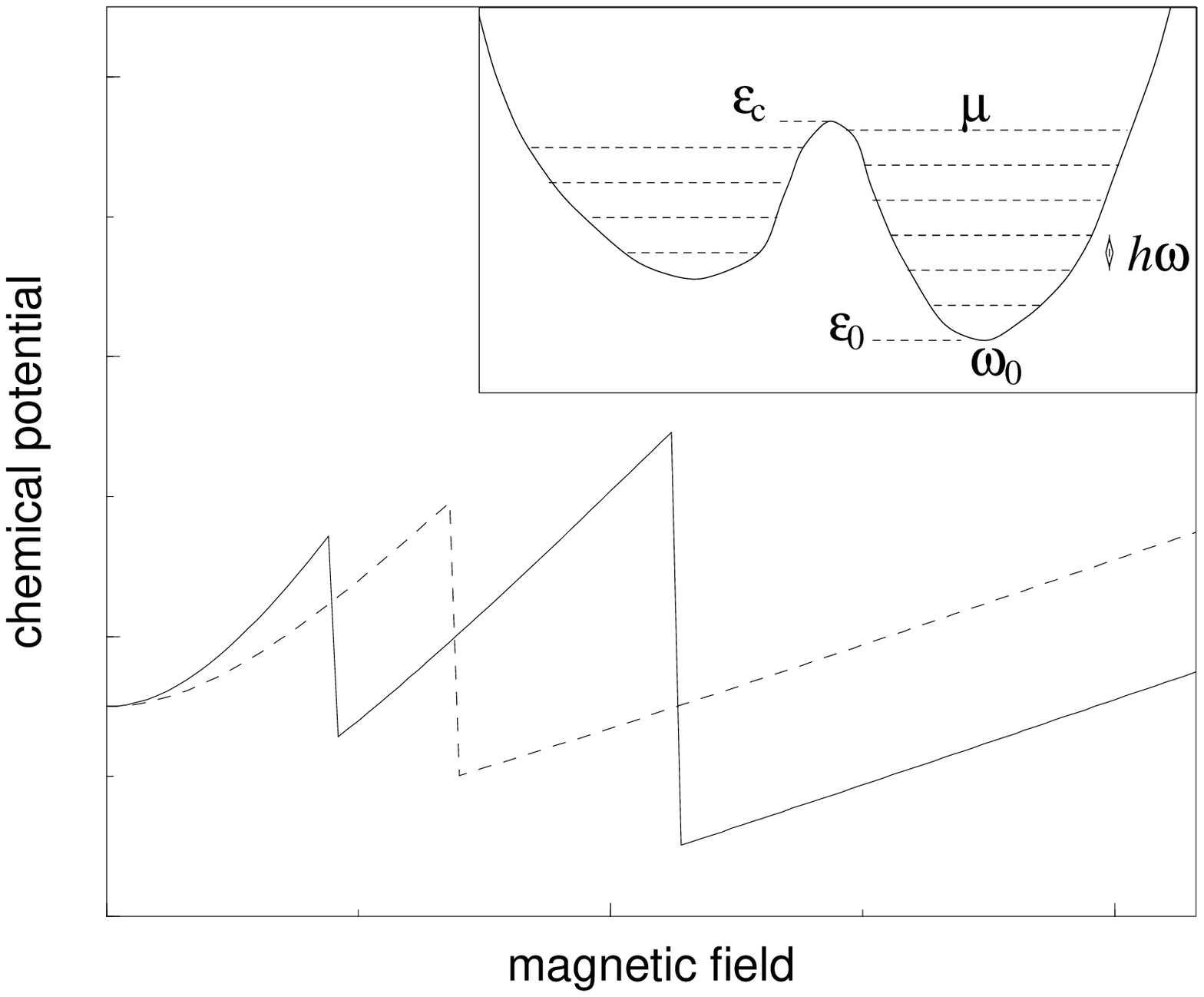}
\end{center}
\vskip -0.5 truecm
\begin{small}
Fig.~1: The magnetic field dependence of the chemical potential in two
different puddles. The chemical potential is stuck to the topmost
occupied Landau level,  which leads to the observed oscillations. The
two puddles differ only in their depth. Inset: the parameters characterizing
a puddle and a neighboring point contact.
\end{small}
\vskip 0.5 truecm

Transport  through a single QPC in perpendicular field  has
been studied experimentally in detail \cite{vanwees}. 
In accordance with the above picture, one finds that the critical
gate voltage (density) oscillates with magnetic field 
due to the depopulation of Landau levels.

The system consists of many puddles, each having its own characteristic
physical parameters. 
Consequently, the integral over the distribution of the
conductances in (\ref{EMTequation}), which depends on the
distribution of the local puddle parameters through the
local chemical potential, $\mu(\mu_0,\e_0,\w_0,B)$, 
has to appropriately performed.
Then Eq.(\ref{linear}) turns into
\begin{eqnarray}
\sigma(B,\mu_0) &=& \sigma_0 \times \left[\overline{\mu}(\mu_0,B)-\mu_c\right]
 ,\ \ \ \ \ 
 \ \ \ \  {\rm with} \nonumber\\
\overline{\mu}(\mu_0,B)&& \!\!= \nonumber\\
\int d\e_0'\,f_{\e_0}(\e_0')\!\!
&&\!\!\!\!\int d\w_0' f_{\w_0}(\w_0')
\ \left[\mu(\mu_0,\e_0,\w_0,B)-\hbar\w/2\right] ,  
\label{averagedmu}
\end{eqnarray}
where $f_{\e_0}$ and $f_{\w_0}$ are the distribution functions of the
respective puddle parameters. In the following these were assumed to
be uniform distributions around characteristic energy scales $\e_0$
and $\w_0$,  respectively,  with respective widths $\Delta\e_0$ and
$\Delta\w_0$. The energy of the lowest Landau level, $\hbar\w/2$, was
subtracted, as it does not contribute to the kinetic energy of the
electrons (or holes) \cite{halperin}. 

\begin{center}
\leavevmode
\epsfxsize=4in
\epsfbox[42 12 606 480]{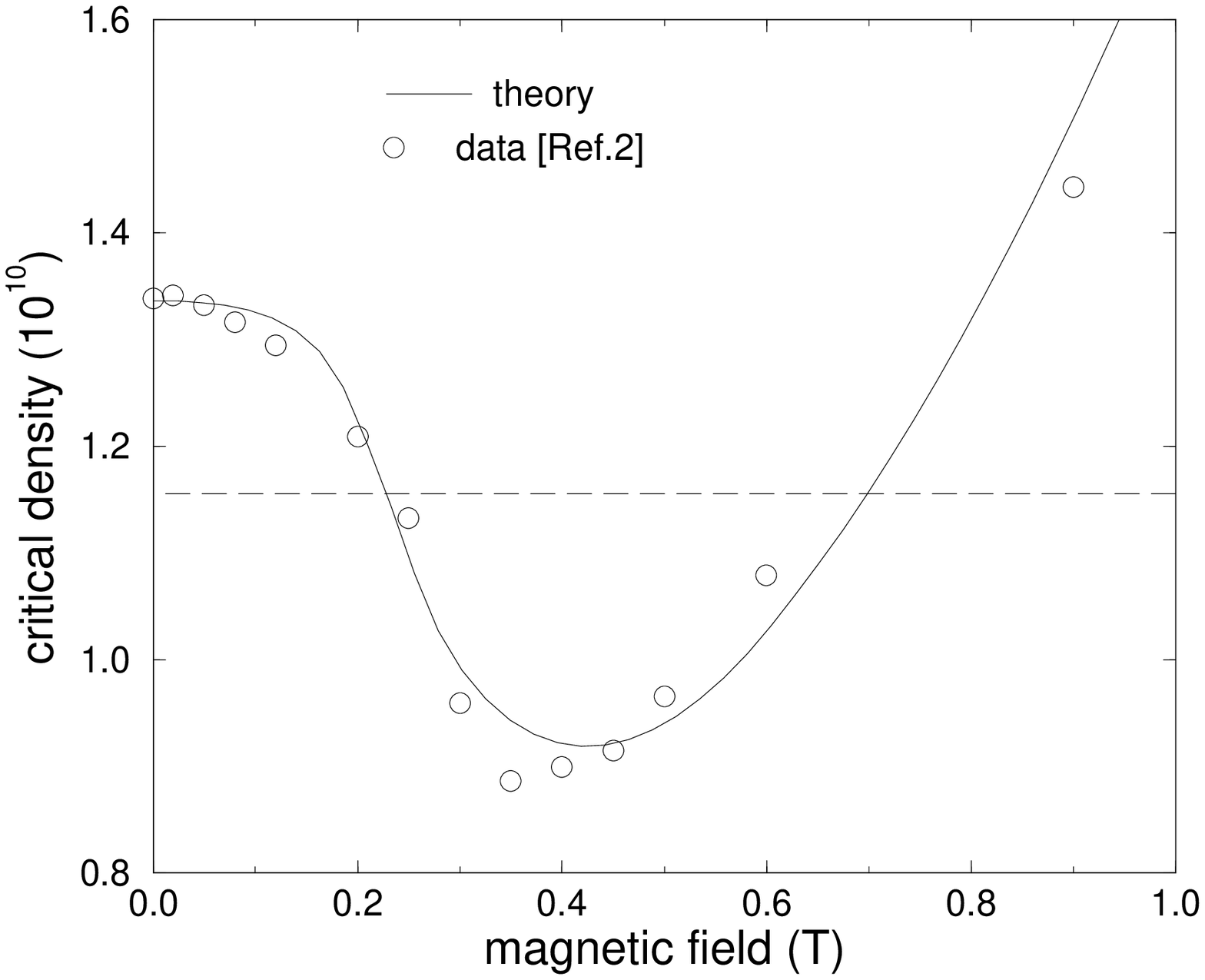}
\end{center}
\vskip -0.5 truecm
\begin{small}
Fig.~2: Comparison of the experimental data \cite{hanein} for the critical
density to the prediction of effective medium theory. The excellent agreement
between theory and experiment allows determination of the puddle potential
distribution (see text). Note that if the magnetic field is changed for
a fixed density (e.g. $n=1.15\times 10^{10}$ along the broken line), there will
 be a finite region of magnetic fields where conductance, in the 
zero-temperature limit, will be non-zero, in contradiction with the
 quantum-phase-transition description of the quantum Hall effect.
\end{small}
\vskip 0.5 truecm

The magnetic field dependence of the 
critical density $\mu_0(B)$ is now determined by the relation
\begin{equation}
\overline{\mu}(\mu_0,B) = \mu_c   .
\label{muc}
\end{equation}

For large fields ($\nu\leq1$) the chemical potential in each puddle
 varies linearly with magnetic field (see Fig.~1).
 Thus it is clear that the averaged 
chemical potential will also vary linearly with field at large fields. As the
field is reduced there will be a discontinuous jump in the chemical potential
of each puddle at it respective integer filling factors.
 One might expect that after averaging
only the largest jump, at $\nu=1$, may survive, and will be replaced by
 a smooth rapid increase. This  has indeed been seen experimentally
 \cite{hanein}. Fig.~2 depicts the experimental data, compared to the 
predictions of the effective-medium theory, Eqs.(\ref{averagedmu},\ref{muc}). 
The parameters $\e_0$ and $\w_0$
determine the value of the magnetic field where the rapid increase in
the critical density occurs (around $\nu=1$), and the saturation value
at zero magnetic field. The width of the puddle potential distribution 
$\Delta\w_0$ determines the rate of increase of the critical density
near $\nu=1$. The excellent agreement between the experimental data and
theory is evident, and allows a determination of the physical characteristics
of the puddle distribution. One finds that the typical confining energy
in the puddle is $0.5$meV, (with about $10\%$ dispersion), leading to
 a typical puddle size of 25 nm.

\begin{center}
\leavevmode
\epsfxsize=3.8in
\epsfbox[42 12 606 480]{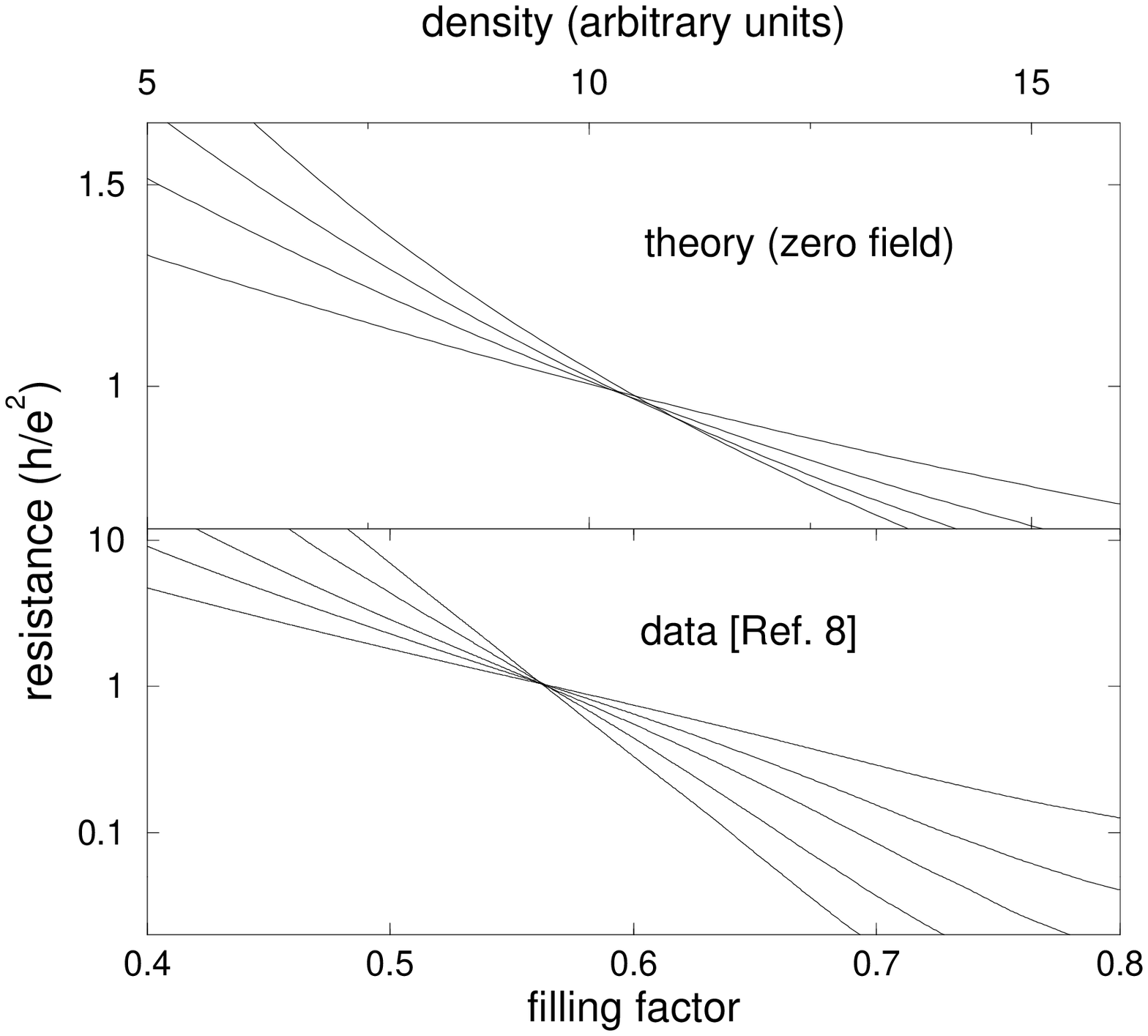}
\end{center}
\vskip -0.5 truecm
\begin{small}
Fig.~3: The exponential dependence of the resistance on density, experiment
\cite{exponent} vs. theory (at zero field). The magnetic field, according
to the theory only shifts the transition. As pointed out in \cite{myperc},
the ``temperature independent point'' in the model has nothing to do with the
critical point, which, in this case, is $n=1.05$, way off the scale. Thus 
all the region depicted in the figure is on the metallic side, which might
explain why the Hall resistance remains quantized, as the density crosses
this ``temperature independent point'' \cite{qinsulator}.
\end{small}
\vskip 0.5 truecm

The experimentally demonstrated connection between the quantum Hall
transition and the zero-field transition indicates that both stem from
the same physical process. The fact that the quantum Hall transition, at
least in these systems, is driven by percolation-like behavior and cannot
be described by the usual theory of quantum phase transition is demonstrated
by the fact that the experimental data in Fig.~2 shows that for as the
magnetic field changes a fixed
density,  $n$ (e.g. $n=1.15\times 10^{10}$ along the broken line in the figure), 
there is a finite (and substantial) range in magnetic fields, where the
zero-temperature limit of the conductance remains nonzero. This is in
 contradiction with the usual theory of quantum phase transition that
predicts, at zero temperature, a single point where transport is 
dissipative. Such data has already been reported in the past \cite{finitesig}, 
and the density and 
temperature dependence of the longitudinal resistance was well 
described by the relation
$\rho_{xx}(n,T)=\rho_1\exp[(n-n_c)/(\alpha T +\beta)]$,  where $n_c$ is the 
critical density and $\rho_1, \alpha$ and $\beta$ some
 constants (note the finite
resistivity when extrapolated to zero temperature). This behavior looks
remarkably like the one reported for the zero-field transition \cite{exponent}, 
$\rho(n,T)=\rho_0 + \rho_1\exp[(n-n_c)/(\alpha T)$.
In order to check this functional dependence, the effective-medium equation
(\ref{EMTequation}) was solved for finite temperatures,  where a uniform
distribution of critical energies $\e_c$ was assumed for simplicity.
The resulting effective resistance is depicted in Fig.~3,  and compared
 to the experimental observation in the quantum Hall regime
 \cite{exponent}. Similarly to the experimental data,  the theory shows
 that there is a wide range in temperature (a high-temperature regime
  in the theory) where the effective resistance
 displays exponential  temperature dependence \cite{scale}. 

The fact that the quantum Hall transition,  at least in some of the
reported observations, stems from the same mechanism that leads to
the metallic behavior in zero magnetic field is further supported by
the fact that  
the 
reported current-voltage duality across the quantum Hall transition 
\cite{duality} was also reported for the zero-field transition \cite{duality0}.
The relevance of percolation to the quantum Hall transition has already been
established experimentally \cite{qhpercolation},  where it was shown that
the transition occurs when the metallic phase percolates through the
system,  and not at a fixed filling factor,  as expected from the 
quantum-phase-transition scenario.

The relevance of the underlying puddle structure and the finite dephasing
rate to these observations in the quantum Hall regime has already
been pointed out and elaborated upon by Shimshoni and coworkers
\cite{efrat}. Also, Pryadko and Auerbach have claimed \cite{dephasing}
that the finite 
quantized value of the Hall conductance reported in the 
insulating regime \cite{qinsulator} can only be explained when finite
dephasing is taken into account, and the Hall resistance
should be infinite in a truly quantum coherent Hall insulator. It 
should be noted though, that as pointed out in \cite{myperc} the identification
of the ``temperature independent point'' with the critical point is
questionable, and within the model it lies well within the metallic phase.
In this picture the fact that the Hall conductance remains
quantized through the ``transition'' is a trivial issue, as this point
is not the true transition point to the insulator. For the parameters 
corresponding to the curves depicted in Fig.~3, the critical point occurs
at $n=1.05$, thus all the points in the figure lie deep in the metallic phase,
and no change in the quantized Hall resistance is expected.

To conclude, the excellent agreement with the experimental data, in addition to 
previously demonstrated quantitative agreement with temperature and
parallel magnetic field data is a further validation of the relevance
of the theory to the description of this phenomenon.
The theory further predicts that when the conductance is measured in
a mesoscopic piece of the system, the conductance will show abrupt
changes, as a function of the density, as point contacts open with
increasing density. Such mesoscopic jumps in the conductance in the
quantum Hall regime have indeed recently observed \cite{jumps}.

The author would like to thank D. Shahar and his group for valuable
discussions and for making their data available to him. This work
  was supported by  the Israel Science Foundation  - Centers of Excellence
 Program,  and by the German Ministry of Science.

\end{multicols}

\begin{references}
\bibitem{abrahamsreview}
For a recent review of experimental data and theoretical approaches,
see E.~Abrahams, S.~V. Kravchenko and M.~P.~Sarachik, cond-mat/0006055.

\bibitem{hanein}
\quote{Y.~Hanein et al.}{Nature}{400}{735 (1999)}

\bibitem{myperc}
Y. Meir, \prl {\bf 83}, 3506 (1999);
\pr {\bf B 61}, 16470 (2000).

\bibitem{kirkpatrick}
S. Kirkpatrick, \prl {\bf 25}, 1722 (1971);
R.~B.~Stinchcombe, {\sl J. Phys.} {\bf C 7}, 179 (1974).

\bibitem{vanwees}
See, e.g., \quote{B. J. van Wees}{\pr}{B43}{12431 (1991)}

\bibitem{halperin}
\quote{H.~A. Fertig and B. I. Haleprin}{\pr}{B 36}{7969 (1987)}

\bibitem{finitesig}
\qquote{M. Hilke et al.}{\pr}{B 56}{R15545 (1997)}
\quote{D. Shahar et al.}{Solid State Comm}{107}{19 (1998)}

\bibitem{exponent}
\qquote{V. M. Pudalov}{JETP Lett.}{66}{175 (1997)}
Y.~Hanein et al., \prl {\bf 80}, 1288 (1998).


\bibitem{scale}
It should be noted, though, that the resistance change in the data is much
larger than that in the theory.

\bibitem{duality}
\qquote{D. Shahar et al.}{Science}{274}{589 (1996)}
\quote{E. Shimshoni et al.}{\pr} {B 55}{13730 (1996)}

\bibitem{duality0}
\quote{D. Simonian,  S. V Kravchenko and M. P. Sarachik}{\pr}{B 55}
{R13421 (1997)}


\bibitem{qhpercolation}
A. A. Shashkin et al.,  \prl {\bf 73},  3141 (1994); V.~T.~Dolgapolov et al., 
JETP Lett. {\bf 62},  162 (1995); I.~V.~Kukuskin et al., \pr {\bf B53},  R13260
(1996).

\bibitem{efrat}
E. Shimshoni and A. Auerbach, \pr {\bf B 55},  9817 (1997);
E. Shimshoni, A. Auerbach and A. Kapitulnik, \prl {\bf 80}, 3352 (1998).
E. Shimsoni,  \pr {\bf B 60},  10691 (1999)

\bibitem{dephasing}
\quote{L.~P.~Pryadko and A.~Auerbach}{\prl}{82}{1253 (1999)}

\bibitem{qinsulator}
\quote{M. Hilke et al.}{Nature}{395}{675 (1998)}

\bibitem{jumps}
D. Shahar, unpublished.

\end{references}
\end{document}